\documentclass[journal]{vgtc}                     

\onlineid{1210}

\vgtccategory{Research}

\vgtcpapertype{theory/model}

\title{Adaptive Assessment of Visualization Literacy}

\author{%
  \authororcid{Yuan Cui}{0000-0002-2681-6441},
  \authororcid{Lily W. Ge}{0000-0003-2350-8686},
  \authororcid{Yiren Ding}{0000-0001-8983-9117},
  \authororcid{Fumeng Yang}{0000-0002-8401-2580},
  \authororcid{Lane Harrison}{0000-0003-3029-2799}, and
  \authororcid{Matthew Kay}{0000-0001-9446-0419} 
}

\authorfooter{
  \item Yuan Cui, Lily W. Ge, Fumeng Yang, and Matthew Kay are with Northwestern University. E-mail: charlescui@u.northwestern.edu, \{wanqian.ge, fy, mjskay\}@northwestern.edu.
  \item Yiren Ding and Lane Harrison are with Worcester Polytechnic Institute. E-mail: \{yding5, ltharrison\}@wpi.edu.
}

\abstract{%
  Visualization literacy is an essential skill for accurately interpreting data to inform critical decisions. Consequently, it is vital to understand the evolution of this ability and devise targeted interventions to enhance it, requiring concise and repeatable assessments of visualization literacy for individuals. However, current assessments, such as the Visualization Literacy Assessment Test (\textsc{vlat}), are time-consuming due to their fixed, lengthy format. To address this limitation, we develop two streamlined computerized adaptive tests (\textsc{cat}s) for visualization literacy, \AVLAT{a-vlat} and \ACALVI{a-calvi}, which measure the same set of skills as their original versions 
in half the number of questions. Specifically, we (1) employ item response theory (\textsc{irt}) and non-psychometric constraints to construct adaptive versions of the assessments, (2) finalize the configurations of adaptation through simulation, (3) refine the composition of test items of \ACALVI{a-calvi} via a qualitative study, and (4) demonstrate the test-retest reliability (\edit{\textsc{icc}}: \AVLAT{0.98} and \ACALVI{0.98}) and convergent validity (\edit{correlation}: \AVLAT{0.81} and \ACALVI{0.66}) of both \textsc{cat}s via four online studies. We discuss practical recommendations for using our \textsc{cat}s and opportunities for further customization to leverage the full potential of adaptive assessments. All supplemental materials are available at \url{https://osf.io/a6258/}. 
}


\teaser{
  \centering
  \includegraphics[width=\linewidth]{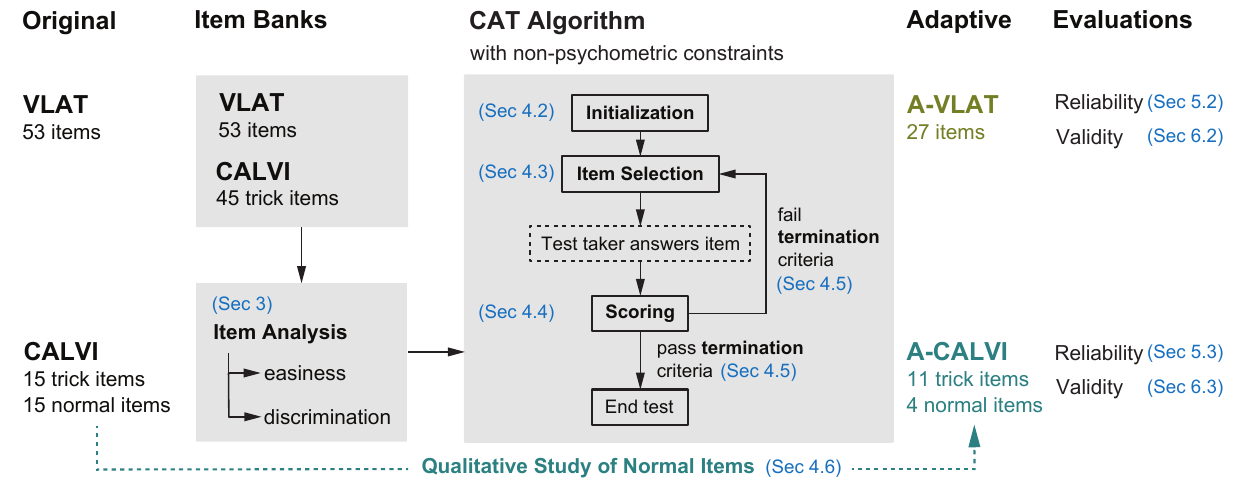}
  \vspace*{-10pt}
  \caption{The process of developing adaptive visualization literacy assessments: \AVLAT{a-vlat} and \ACALVI{a-calvi}. We start with the item banks of  \textsc{vlat} and \textsc{calvi} and use the parameters from item analysis to construct the \textsc{cat} algorithms for the adaptive assessments. We then evaluate their validity and reliability via four online studies. The annotations in blue are the components' corresponding sections.
  }
  \label{fig:teaser}
}

\graphicspath{{figs/}{figures/}{pictures/}{images/}{./}} 

\usepackage{tabu}                      
\usepackage{booktabs}                  
\usepackage{lipsum}                    
\usepackage{mwe}                       
\usepackage{mathptmx}                  
\usepackage{xfrac}

\usepackage{amsmath}
\usepackage{mathtools}
\usepackage[hang,flushmargin]{footmisc} 
\usepackage{xurl}
\usepackage[defaultlines=1,all]{nowidow}

\newif\ifnotes
\notestrue 

\newcommand{\AVLAT}[1]{\textcolor{olive}{\textsc{#1}}}
\newcommand{\ACALVI}[1]{\textcolor{teal}{\textsc{#1}}}
\newcommand{\code}[1]{\ttfamily{\fontsize{7.75}{7.75}\selectfont{#1}}\normalfont}
 
  \makeatletter
\renewcommand{\paragraph}{%
  \@startsection{paragraph}{4}%
  {\z@}{.5ex \@plus .5ex \@minus .4ex}{-.5em}%
  {\fontfamily{qhv}\fontsize{8.5}{8.5}\selectfont\textbf}%
}
\makeatother  

\newif\ifedited
\editedfalse 
\usepackage{soul}
\usepackage[dvipsnames]{xcolor}
\definecolor{graycolor}{RGB}{180, 180, 180}
\definecolor{editcolor}{RGB}{227, 26, 16}
\newcommand{\edit}[1]{\ifedited{\leavevmode\color{editcolor}{#1}}\else{#1}\fi}

\definecolor{mkcolor}{HTML}{e32d97}
\definecolor{fycolor}{HTML}{9760d6}
\definecolor{charlescolor}{HTML}{0b9e41}
\definecolor{yirencolor}{HTML}{f27013}
\definecolor{lilycolor}{HTML}{e8b94d}
\definecolor{lanecolor}{HTML}{4e81d4}
\definecolor{mandicolor}{HTML}{f030e6}

\begin{document}

\setlength{\abovedisplayskip}{5pt}
\setlength{\belowdisplayskip}{5pt}
\setlength{\abovedisplayshortskip}{5pt}
\setlength{\belowdisplayshortskip}{5pt}
\setlength{\belowcaptionskip}{-5pt}
\setlength{\abovecaptionskip}{2pt}

\maketitle

\section{Introduction} 
Visualization literacy---an individual's ability to understand and interpret visualizations---can significantly impact data-driven decisions. 
People may rely on data visualizations showing the spread of infectious diseases to make personal health decisions, to decide between treatment options in medical settings, to make financial decisions, or to engage with social or political topics.
Inaccurate interpretations of such visualizations may lead to faulty reasoning and decisions, as well as harmful outcomes. 


In the study of visualization literacy, there is a persistent need for accurate, reliable, and timely ways to assess people's ability.
For example, visualization researchers may want to  
track the progress of this ability over time, or to empirically evaluate interventions designed to enhance a target group's ability to interpret visualizations. 
Such efforts require concise, quantitative assessments of visualization literacy that can be administered to individuals multiple times, allowing for the measurement of skill development and the evaluation of intervention efficacy through pre- and post-testing.

While researchers have devised assessments to measure both basic visualization literacy skills \cite{Lee2017VLAT, boy2014principled}
and the critical thinking ability to detect visualization misinformation \cite{CALVI}, these tests can be time-consuming.
To address this,
we apply \textit{computerized adaptive testing} (\textsc{cat}) to visualization literacy. \textsc{cat}'s core principle is to adaptively select test items for a test taker based on their performance on previously-answered items, and it can provide a precise estimate of the test-taker's abilities with fewer items. \textsc{cat} has been widely applied in healthcare science \cite{gibbons2016electronic, gibbons2014developmentcatanx, walter2007developmentanxcat, gibbons2019without} and used in practice by educational agencies, such as the Educational Testing Service (\textsc{ets}) and College Board, to create large-scale standardized tests \cite{greadaptive, rudner2009gmat, satgodigital}. However, visualization literacy assessments have yet to take advantage of the benefits of adaptive testing\looseness=-1.


In this paper, we adopt the \textsc{cat} development framework~\cite{thompson2011framework} 
to create two short, adaptive visualization literacy tests: \AVLAT{a-vlat} and \ACALVI{a-calvi}, which are built upon the existing static assessments \textsc{vlat}~\cite{Lee2017VLAT} and \textsc{calvi}~\cite{CALVI}. 
First, we compute \textit{item parameters} (how easy items are and how well items in the bank separate test takers of different abilities) 
with Item Response Theory (\textsc{irt}). Using the item parameters, we construct adaptive algorithms to select items for test-takers. We include non-psychometric constraints in our algorithms to ensure the tests
contain balanced content; i.e., to ensure \AVLAT{a-vlat} covers all 12 chart types and 8 tasks from \textsc{vlat} and \ACALVI{a-calvi} covers all 11 misleaders from \textsc{calvi}.
We also conduct a qualitative study to refine the composition of items in \ACALVI{a-calvi}, reducing its length by a further 11 items. 
Ultimately, we contribute:
\begin{enumerate}
    \item   A demonstration of the refinement of visualization literacy assessments through adaptive testing, including the incorporation of non-psychometric features of existing assessments.
\item Two valid and reliable adaptive visualization literacy tests, \AVLAT{a-vlat} (27 items) and \ACALVI{a-calvi} (15 items), which are half the length of their non-adaptive counterparts.
\item Evidence from four online studies demonstrating the test-retest reliability (\edit{\textsc{icc}}: \AVLAT{0.98} and \ACALVI{0.98}) and convergent validity (\edit{correlation}: \AVLAT{0.81} and \ACALVI{0.66}) of these tests. 

\end{enumerate}

In addition, we discuss how cumulative results from visualization literacy studies can better inform our understanding of the relationship between the constructs measured by literacy assessments (e.g., what correlations between chart types might tell us about how people understand visualizations). We also 
provide recommendations for using and customizing visualization literacy assessments based on test administrators' needs. Our work just scratches the surface of the potential for adaptive testing in visualization; we believe it offers a path to shorter, repeatable, and more reliable assessment of visualization literacy.




\section{Background}
\subsection{Visualization Literacy}
\label{subsec:vislit}

Within the visualization community, many researchers have studied visualization literacy through the development of assessments \cite{boy2014principled,Lee2017VLAT,CALVI} \edit{and frameworks \cite{DVL}}. Boy et al. used Item Response Theory (\textsc{irt}) to generate a set of tests that aims to assess people's ability to interpret line charts, bar charts, and scatterplots \cite{boy2014principled}. Similarly, to test people's ability to interpret visually represented data, Lee et al. developed a Visualization Literacy Assessment Test (\textsc{vlat}) that contains 53 multiple-choice items \cite{Lee2017VLAT}. Later, considering the complexity of visualization literacy as a construct, Ge et al. expanded on prior definitions to incorporate the ability to identify and reason about visualization misinformation and developed a Critical Thinking Assessment for Literacy in Visualizations (\textsc{calvi}), which has a bank of 45 items \cite{CALVI}. We rely heavily on both \textsc{vlat} and \textsc{calvi} \edit{to develop} our adaptive tests, \AVLAT{a-vlat} and \ACALVI{a-calvi}, so we explain \textsc{vlat} and \textsc{calvi} in more detail below.

\begin{figure*}[tbp]
  \centering
  \includegraphics[scale=0.80]{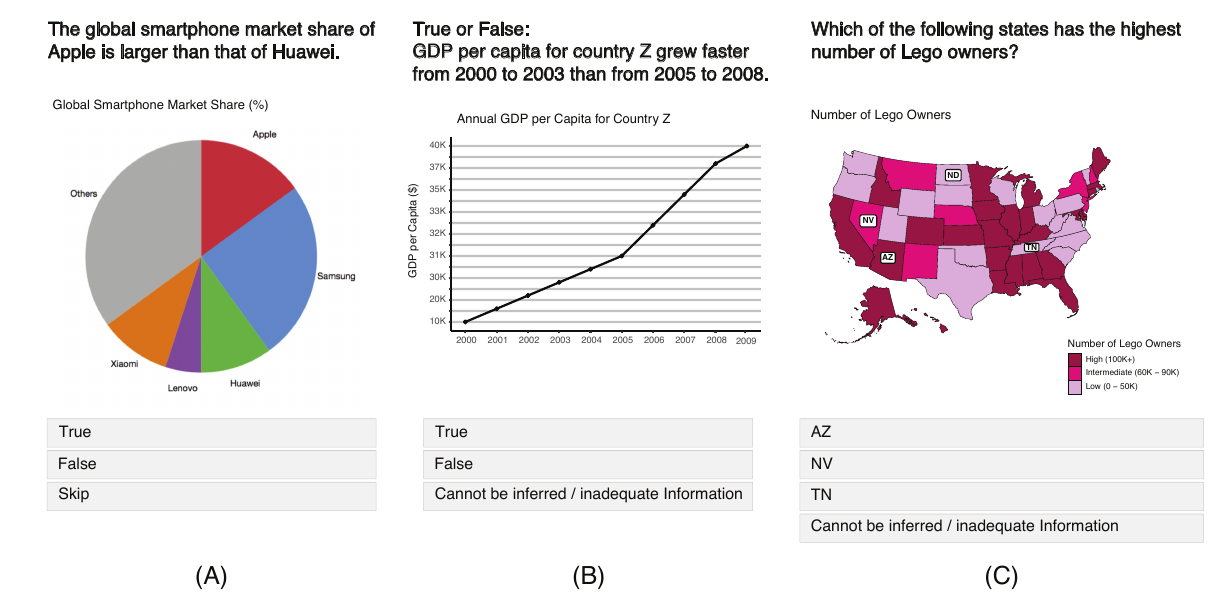}
  \caption{Example items from \textsc{vlat} and \textsc{calvi}: (A) is an item from \textsc{vlat}; (B) is a trick item from \textsc{calvi}; (C) is a normal item from \textsc{calvi}. 
  }
\label{fig:itemexamples}
\end{figure*}

\paragraph{VLAT} The items in \textsc{vlat} were generated from 12 chart types with 8 tasks such as \textit{retrieve value} and \textit{make comparisons}, and each test taker is expected to take all 53 items \cite{Lee2017VLAT}. For example, \cref{fig:itemexamples}.A is a \textsc{vlat} item in a pie chart that asks viewers to make comparisons. Each item in \textsc{vlat} has an item difficulty and item discrimination index obtained from Classical Test Theory (\textsc{ctt}) analysis \cite{Lee2017VLAT}. 

\paragraph{CALVI} The items in \textsc{calvi} were generated from 11 misleaders (i.e., ways a chart can lead to conclusions not supported by the data) and 9 chart types \cite{CALVI}. \textsc{calvi} consists of 15 \textit{trick} items, which are items whose visualizations contain a misleader, and 15 \textit{normal} items (i.e., items based on well-formed visualizations)---\textsc{calvi} includes normal items because people may encounter misleading visualizations mixed in with well-formed visualizations in their daily lives. It is worth noting that only test takers' performance on the trick items are used to estimate their abilities to detect visualization misinformation.
The 15 trick items are selected to cover all 11 misleaders from \textsc{calvi}'s bank of 45 total trick items \cite{CALVI}. For example, \cref{fig:itemexamples}.B is a trick item with misleader \textit{Manipulation of Scales - Inappropriate Use of Scale Functions}, where the number labels on the $y$-axis do not match the spacings of the tick marks; and \cref{fig:itemexamples}.C is a normal item. The trick items in \textsc{calvi} each have an item easiness and an item discrimination parameter estimated using a 2-parameter \textsc{irt} model \cite{CALVI}.

Besides the assessments \edit{above}, related fields such as statistical literacy also investigated graph interpretation abilities, and measurements were developed to assess graph comprehension. Specifically, studying how students interpreted graphical representations of distributions, DelMas et al. developed assessment items for the ability to interpret distributions~\cite{delmas2005using}.
Other researchers have also developed instruments 
that target younger audiences~\cite{Peterman2015instrument}, and some focused more on the ability to critically interpret graphs, which is related to statistical literacy~\cite{aoyama2003graph}\looseness=-1.

Although existing assessments provide measurements for visualization literacy and related abilities, we reiterate that they can be time consuming and inflexible due to their fixed, lengthy format.
Next, we give an overview of computerized adaptive testing, which reduces the length of tests while maintaining similar measurement precision. 




\subsection{Computerized Adaptive Testing (\textsc{CAT})}

Computerized adaptive testing (\textsc{cat}) is a form of computer-based test where the next item or the next set of items a test-taker sees depends on their performance on the previous items. The idea is to adaptively select items to tailor to a test-taker's ability in order to gain more information about them: for example, if someone performs poorly on difficult items, they will then be presented with an easier item. \textsc{cat} has many advantages over traditional static assessments where everyone receives the same items: it can achieve similar measurement precision with fewer items, and it may better motivate test takers because the items are more appropriate for their ability level \cite{rezaie2015computer}.


\textsc{cat} has been broadly adopted in healthcare research. Clinical researchers have developed various \textsc{cat}s to measure patient characteristics, such as quality of life \cite{gibbons2016electronic}, anxiety \cite{gibbons2014developmentcatanx, walter2007developmentanxcat}, and mental health disorders \cite{gibbons2019without}. \textsc{cat} has also been used to measure various forms of literacy, such as English literacy \cite{foorman2016englishlit}, health literacy \cite{kandula2011healthlit}, and math knowledge of university students \cite{ghio2022mathlit}.

\textsc{cat} has also been widely applied to many large-scale standardized tests in practice, such as the Graduate Record Examinations (\textsc{gre}) \cite{greadaptive} and the Graduate Management Admission Test (\textsc{gmat}) \cite{rudner2009gmat}, and continues to grow in popularity: the SAT, a long-established college admissions test in the U.S., is planning to become digital and adopt adaptive testing starting in 2023 and 2024 \cite{satgodigital}.

Despite the prevalent prior applications and potential benefits of \textsc{cat}, it has yet to be applied towards visualization literacy assessments. 
We address this gap and explore the benefits of \textsc{cat} in visualization literacy, developing shorter, adaptive assessments, along with a systematic \textsc{cat} development procedure across two existing visualization literacy tests. 





\subsection{Procedure of \textsc{CAT} Development}\label{subsec:procedureofcatdev}

Although existing measurement methods are not adaptive, as explained in \cref{subsec:vislit}, they provide a valuable foundation for the development of \textsc{cat}, which requires item parameters from \textsc{irt} item analysis.
We extend 
and adopt the framework proposed by Thompson and Weiss \cite{thompson2011framework} to develop adaptive tests. Below, we outline steps from this framework that were instrumental to developing our test.

First, test developers need to conduct \textbf{item analysis} of the item bank using test tryout data (i.e., data collected on participants answering items before the test is deployed in practice). In this stage, \textsc{irt} item analysis is applied on test tryout data to compute the item easiness (i.e., how easy an item is) and item discrimination (i.e.,  how well an item differentiates test takers of different levels of ability) parameters
of the items in the bank. After the test developers obtain these parameters for the items, they can proceed to the \textbf{construction of \textsc{cat} algorithms}. A \textsc{cat} algorithm consists of four main components: 
\begin{enumerate}
    \item Initialization: initialize an estimate of the test-taker's ability;
    \item Item selection: select the next item for the test-taker based on their current estimated ability (score); 
    \item Scoring: compute the score for a test-taker after  each item;
    \item Termination: end the test when the termination criterion is met.
\end{enumerate}
To build a \textsc{cat} algorithm, test developers use the test tryout data, the item parameters, and simulation studies to determine the configurations of these main components and demonstrate the measurement precision of the \textsc{cat}.
Although the framework thus far does not outline steps specifically for establishing reliability and validity, in the development of tests in general, they should be assessed \cite{psychTestingBook}. 
Therefore, for both \textsc{cat}s we also evaluate the \textbf{reliability} (i.e., examining the extent to which the test produces consistent and stable results) and \textbf{validity} (i.e., ensuring that the test measures the ability that it is supposed to measure) \cite{psychTestingBook}.

\section{Item Analysis}\label{sec:itemanalysis}
As outlined in \cref{subsec:procedureofcatdev}, the first stage of \textsc{cat} development
is item analysis. 
In this section, we describe the \textsc{irt} model used for item analysis to obtain parameters for both item easiness and discrimination. 
These item parameters are necessary for constructing the \textsc{cat} algorithms, to be detailed in the following sections. 
Additional results from the item parameter analysis are available in supplemental materials.
\subsection{2-Parameter \textsc{IRT} Model}
\textsc{irt} mathematically describes the relationship of a test taker's ability, item parameters, and the probability of them answering a particular item correctly. We use the 2-parameter \textsc{irt} model to infer the item easiness and discrimination parameters of all \textsc{vlat} and \textsc{calvi} items. The item response function of the 2-parameter \textsc{irt} model (\Cref{eq:1}) describes the relationship between the probability of a test taker with ability $\theta$ answering item $i$ correctly (left-hand side) and the easiness $b_i$ and discrimination $a_i$ of item $i$ (right-hand side). This function will be used in conjunction with test tryout data to compute the item parameters (i.e., $a_i$ and $b_i$) for \textsc{vlat} and \textsc{calvi} items, and it will also be used in \cref{sec:constructionofcatalg} to define and compute other key quantities (e.g., item information, standard error) in adaptive testing.


\begin{equation}
  p_i(\theta) = \frac{1}{1 + \exp(-a_i(\theta + b_i))} \label{eq:1}
\end{equation}

\subsection{Parameters of \textsc{VLAT} and \textsc{CALVI} Items}
The developers of \textsc{calvi} collected test tryout data and used the 2-parameter Bayesian \textsc{irt} model to compute the item parameters of each item \cite{CALVI}. Therefore, we can directly use these item parameters to adaptively select the next item for \ACALVI{a-calvi}.

Because item analysis for \textsc{vlat} was conducted with \textsc{ctt} instead of \textsc{irt}, 
we apply \textsc{irt} analysis on \textsc{vlat} items using their test tryout data.\footnote{\edit{This} data was obtained from the authors of \textsc{vlat} by personal communication.}
We use a similar 2-parameter Bayesian \textsc{irt} model as the one used for \textsc{calvi} with adjustments to address the following difference between the two tests: \textsc{calvi} has two types of items: normal items and trick items, and only the trick items are associated with test takers' abilities $\theta$. 
Whereas in \textsc{vlat}, a person's performance on every item counts. All other specifications are the same as the model from \textsc{calvi}, provided in supplemental materials, along with the item parameters of the \textsc{vlat} and \textsc{calvi} items.




\section{Construction of \AVLAT{A-VLAT} and \ACALVI{A-CALVI} Algorithms}\label{sec:constructionofcatalg}

\begin{figure}[ht]
  \centering 
  \includegraphics[width=1.0\columnwidth]{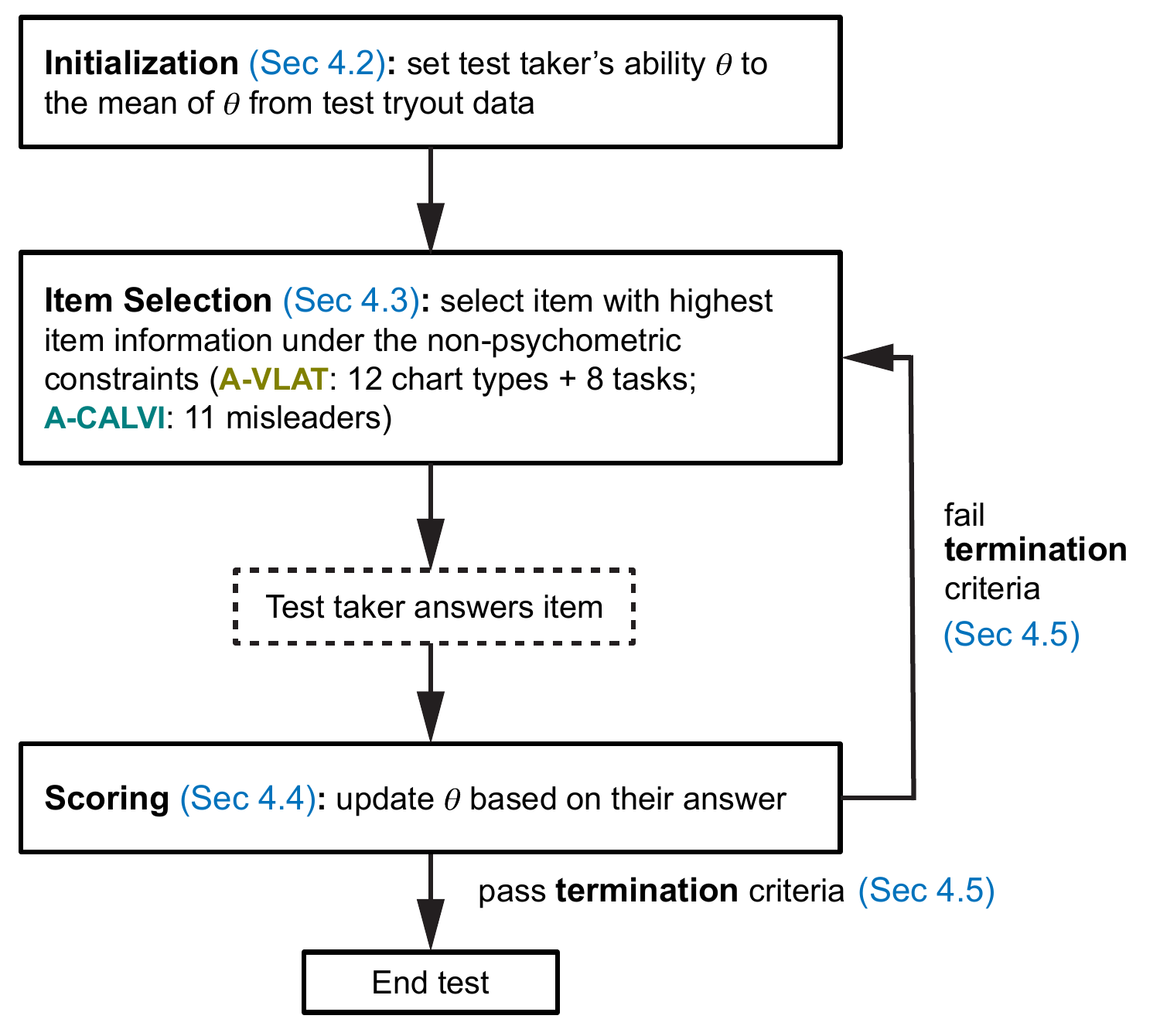}
    \vspace*{-5pt}
  \caption{%
  	The components and flow of our \textsc{cat} algorithm: the algorithm begins by initializing ability $\theta$ for a test taker, which is then used to select the next item. The test taker then answers the item, and their $\theta$ gets updated based on the answer. If the termination criterion is satisfied, then the test ends; if not, the updated $\theta$ is used to select the next item and this process repeats. 
   %
  }
      \vspace*{-5pt}
  \label{fig:catalg}
\end{figure}

To construct \AVLAT{a-vlat} and \ACALVI{a-calvi}, we use test tryout data, item parameters, and simulation studies to determine the configurations of \AVLAT{a-vlat} and \ACALVI{a-calvi}: \textbf{(1)} \textbf{initialization} (\cref{subsec:initialization}), \textbf{(2)} \textbf{item selection} (\cref{subsec:itemselection}), \textbf{(3)} \textbf{scoring} (\cref{subsec:scoring}), and \textbf{(4)} \textbf{termination} (\cref{subsec:termination}). In addition, we conduct a qualitative study for \ACALVI{a-calvi} to determine how to balance items across the misleading / \edit{normal} categories (\cref{subsec:normal_items_study}). \Cref{fig:catalg} presents these components and the flow of our \textsc{cat} algorithm.

\subsection{Key Quantities in \textsc{IRT}-based \textsc{CAT}}\label{subsec:theoryirtcat}
The idea of \textsc{cat} is to iteratively give a test taker an item that is best suited to their ability and iteratively update their estimated ability. 
In \textsc{irt}-based \textsc{cat}, an item that is best suited to a test taker with ability $\theta$ is the item that provides the most information about that test taker; this quantity is called \textit{item information}. \Cref{eq:iteminfo} provides the mathematical definition 
of item information in the 2-parameter model~\cite{baker2001basics}: on the \edit{left-hand} side, $I_i(\theta)$ is the item information from test taker with ability $\theta$ answering item $i$; on the \edit{right-hand} side, $a_i$ is the item discrimination parameter of item $i$, and $p_i(\theta)$ is the probability of person with ability $\theta$ answering item $i$ correctly (defined in \cref{eq:1}). 
\begin{equation}
  I_i(\theta) = a_i^2 p_i(\theta)(1-p_i(\theta)) \label{eq:iteminfo} 
\end{equation}
Notice that $I_i(\theta)$ is maximized when $p_i(\theta) = 0.5$. This means that an item provides the most information about a test taker with ability $\theta$ when \edit{we are most uncertain} about whether the test taker will get this item right.\footnote{\edit{This should not be confused with the test taker being most uncertain about which answer is correct (i.e. randomly picking an option). For instance, if a test taker picks randomly in a 3-option item, $P(\textrm{correct})$ would be $\sfrac{1}{3}$ and $P(\textrm{incorrect})$ would be $\sfrac{2}{3}$---so we would be more certain that they would answer incorrectly than correctly. Uncertainty is maximized at $\sfrac{1}{2}$, not $\sfrac{1}{3}$.}
}
Also, the greater the item discrimination $a_i$ (an item’s ability to differentiate test takers of different levels of ability), the greater the item information. 

Under the context of item information, the item that is best suited to a test taker with ability $\theta$ is the item with the highest information for $\theta$ in the item bank. This will be used in item selection (\cref{subsec:itemselection}). 

Given the definition of item information, we can also define the \textit{test information}~\cite{baker2001basics} as the sum of the item information of its items: 
\begin{equation}
  I(\theta) = \sum I_{i}(\theta)
  \label{eq:iteminfototal}
\end{equation}
Another use of item information is to quantify the measurement precision of the test. The \textit{standard error}~\cite{baker2001basics} is defined as
\begin{equation}
  SE(\theta) = \frac{1}{\sqrt{I(\theta)}}.
  \label{eq:irtse}
\end{equation}
\textbf{The larger the test information, the smaller the standard error, and the more precise the measurement of the test on ability $\theta$.} The standard error will be used in the simulation studies to construct a metric \edit{for comparing} the measurement precision of our adaptive tests with their original static counterparts, and this metric will provide evidence for selecting appropriate lengths for the adaptive tests (\cref{subsec:termination}).

\subsection{Initialization}\label{subsec:initialization}
As shown in \cref{fig:catalg}, the first step of the \textsc{cat} algorithm is to initialize $\theta$ for a test taker, which can then be used for item selection. 
After examining the test tryout data of both \textsc{vlat} and \textsc{calvi}, we find that the distributions of $\theta$ are approximately normal for both tests.
Thus, most people's $\theta$'s are close to the mean of the distribution. 
Therefore, we initialize the $\theta$ of both adaptive tests using the mean of the $\theta$'s in respective test tryout data.
We explore the implications and possibilities of different methods for initializing $\theta$ in the discussion section.


\subsection{Item Selection}\label{subsec:itemselection}
The next item in \AVLAT{a-vlat} or \ACALVI{a-calvi} will be the item that yields the highest item information for the test taker's current $\theta$ estimate. However, simply selecting without constraints could risk not covering the important aspects of the original test, which is a salient trade-off of shorter adaptive tests. This is referred to as \textit{content balancing} \cite{segall2005computerized, wainer2000catprimer}.


To ensure content balancing, we impose non-psychometric constraints to the item selection process such that \AVLAT{a-vlat} covers all 12 chart types and 8 tasks of \textsc{vlat} and \ACALVI{a-calvi} covers all 11 misleaders of \textsc{calvi}. We refer to them as \textit{non-psychometric features}.\footnote{The list of non-psychometric features is provided in supplemental materials.}

This is achieved by the following logic: we initialize the list of non-psychometric features to contain all features needed to be covered. 
If the remaining number of items in the test is \textbf{greater than the length of this list}, then the item with the highest information is selected from the bank, and its associated non-psychometric features are removed from the list. If the remaining number of items in the test is \textbf{equal to the length of this list}, the highest-information item with at least one of the remaining features is selected, and its associated non-psychometric features are removed from the list.  






\subsection{Scoring}\label{subsec:scoring}
In adaptive testing, \edit{the $\theta$ (ability) estimate} needs to be updated every time a test taker answers an item (shown in \cref{fig:catalg}). 
\edit{We initialize} \edit{$\theta$} to a prior distribution \edit{based on} the test tryout data from participants taking \textsc{vlat} and \textsc{calvi}.  
Then, with each answer, we perform a Bayesian update to adjust \edit{the distribution of the $\theta$ estimate}. 
\edit{We compute} the final score of a test taker by taking the mean of the distribution of \edit{the $\theta$ estimate} at the end of this iterative process.

\subsection{Termination}
\label{subsec:termination}
We develop a fixed-length \textsc{cat} for \AVLAT{a-vlat} and \ACALVI{a-calvi}. 
While \textsc{cat}s can be variable-length or fixed-length, variable-length tests can be difficult to implement in formal learning spaces with fixed time constraints, and require additional considerations and procedures to develop.
We determine the lengths using simulations\footnote{The code and data for simulations can be found in supplemental materials.} by comparing the standard error of each person in fixed-length \textsc{cat} of various lengths to that of the original static version of the test. In \cref{subsubsec: vlatcatsim} and \cref{subsubsec:simcalvi}, we present  details of the simulation studies resulting in \textbf{27 items for \AVLAT{a-vlat}} and \textbf{ 11 trick items for \ACALVI{a-calvi}}.

Because we select the original static tests as the baseline for both \AVLAT{a-vlat} and \ACALVI{a-calvi}, we define the following metric:

\begin{equation}
    \textrm{relative difference in }SE(\theta) = \frac{SE_\textrm{A}(\theta) - SE_\textrm{O}(\theta)}{SE_\textrm{O}(\theta)},
\end{equation}
where $SE_\textrm{A}(\theta)$ is the standard error for $\theta$ in the adaptive test and $SE_\textrm{O}(\theta)$ is the standard error for $\theta$ in the original test. This metric is a measure of relative precision of the adaptive test compared to the original version; the smaller $\textrm{relative difference in }SE(\theta)$ is, the more precise the adaptive test compared to the original version. For example, a \textrm{relative difference in }$SE(\theta)$ of 0.30 implies that for this $\theta$, the standard error of the adaptive test is within 30\% of the standard error of the baseline. 
Similarly, a \textrm{relative difference in }$SE(\theta)$ of 0 means that the adaptive and original versions are equally precise. Note that if this quantity is positive, the adaptive test has a larger standard error than the baseline for $\theta$. 
If the quantity is negative, then the adaptive test has a smaller standard error than the baseline for $\theta$. 



\subsubsection{\textbf{Simulation for \AVLAT{A-VLAT}}}\label{subsubsec: vlatcatsim}
In this simulation study, we compare the measurement precision of \AVLAT{a-vlat} with \textsc{vlat} and provide reasoning for why we select 27 items as the length for \AVLAT{a-vlat}.


\paragraph{Data} The item analysis of \textsc{vlat} test tryout data on 191 participants yielded $\theta$ estimates for all 191 participants. We fit a normal distribution for $\theta$ using this data and generate a sample of 500 simulated persons (i.e., $\theta$'s).
We treat the simulated $\theta$'s as the ground truth latent abilities for the simulated persons \edit{and refer to them as \textbf{true $\theta$'s}} below.


\paragraph{Method} Because our goal is to find a reasonable fixed length for \AVLAT{a-vlat}, we build a simulation that realizes the 500 simulated persons' test results under \AVLAT{a-vlat} of different lengths from 19 items\footnote{The non-psychometric constraint of \AVLAT{a-vlat} dictates that it must cover all 12 chart types and 8 tasks. Since any item is associated with a chart type and a task, \AVLAT{a-vlat} needs to have at least 19 items to satisfy the non-psychometric constraint in the worst case.} to 53 items and the original version of \textsc{vlat}, which has a length of 53. We do so by
\edit{(1) plugging in the true $\theta$ and the item parameters into \cref{eq:1} to compute the probability of correctness for each simulated person, (2) simulating the binary correctness outcome with that probability, and (3) computing the mean of the posterior distribution of the $\theta$ estimate as the final score for each person as described in \cref{subsec:scoring}.
}

From here, we compute the $\textrm{relative difference in }SE(\theta)$ for all 500 simulated persons.

\begin{figure}[t]
  \centering 
  \includegraphics[width=\columnwidth]{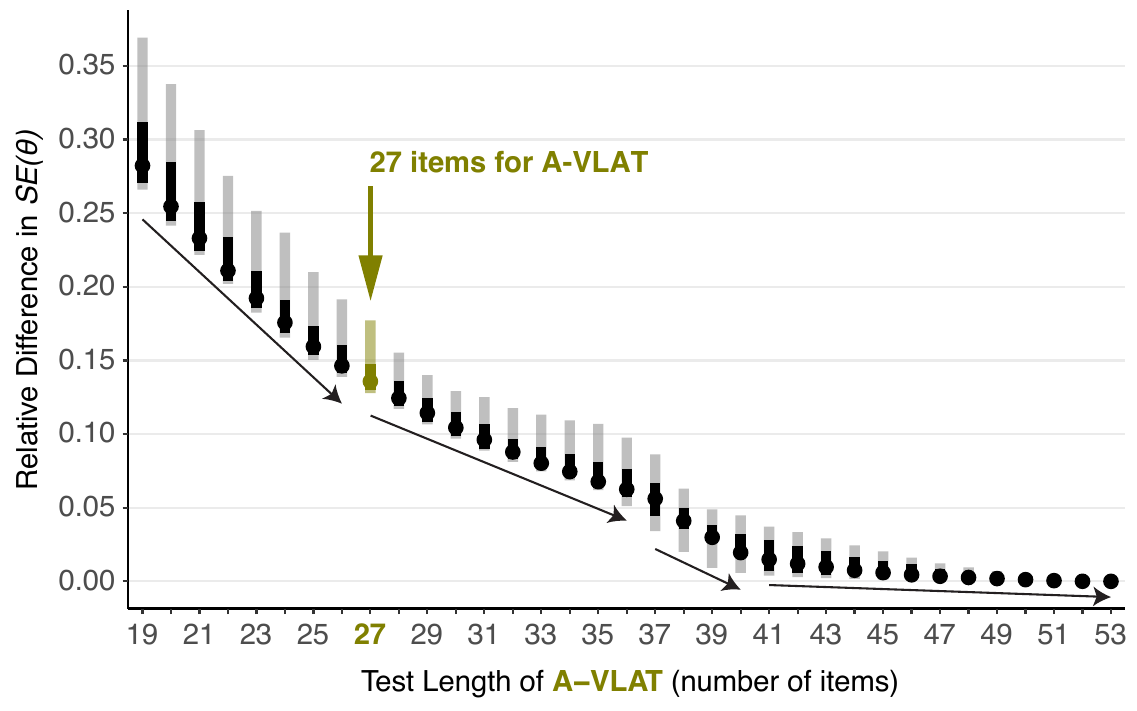}
  \caption{%
  	The distribution of $\textrm{relative difference in }SE(\theta)$ from length 19 to 53 for \AVLAT{a-vlat}: the dot is the median, the lighter bar is where 95\% of the observations fall, and the darker bar is where 66\% of the observations fall. The trend has approximately 4 segments as indicated by the arrows. \edit{We select 27 as the length of \AVLAT{a-vlat.}}
  }
    \vspace*{-10pt}
  \label{fig:simresultsvlat}
\end{figure}

\paragraph{Results} \Cref{fig:simresultsvlat} shows the 500 simulated persons' distribution of $\textrm{relative difference in }SE(\theta)$ for \AVLAT{a-vlat} of length 19 to 53. Upon inspecting the trend, we find that this chart has approximately four segments: $\textrm{relative difference in }SE(\theta)$ experienced the sharpest drop from length 19 to length 26, a milder decrease from 27 to 36, another steep yet short downfall from 37 to 40, and finally plateaus after 41. Note that it never goes below 0 because the baseline uses all 53 items, and adaptively selecting 53 items would yield the same result. 

After examining \cref{fig:simresultsvlat}, we select 27 as the length of \AVLAT{a-vlat}.
\edit{This reduces the length of the original test by half, and the median and most observations of $\textrm{relative difference in }SE(\theta)$ are below 0.15, a reasonable loss of measurement precision in exchange for halving the test. It should be noted that there is no gold standard for selecting the perfect cutoff in the literature. Test administrators can use \cref{fig:simresultsvlat} to help customize the length based on their needs and constraints. We will discuss customization of our adaptive tests in \cref{sec:cats:customization}.} 


\subsubsection{\textbf{Simulation for \ACALVI{A-CALVI}}}\label{subsubsec:simcalvi}

In this simulation study, we compare the measurement precision of \ACALVI{a-calvi} with \textsc{calvi} and provide reasoning for why we select 11 as the number of trick items for \ACALVI{a-calvi}.

\paragraph{Data} Similar to the \AVLAT{a-vlat} simulation, we used the $\theta$ estimates for all 497 participants from \textsc{calvi} test tryout to generate a sample of 500 simulated persons. We treat all simulated $\theta$'s as the ground truth latent abilities for the simulated persons.  

\paragraph{Method} The same method in \cref{subsubsec: vlatcatsim} was applied. we obtained the $\textrm{relative difference in }SE(\theta)$ for all simulated persons.\footnote{Similar to \AVLAT{a-vlat}, \ACALVI{a-calvi} needs at least 11 items to cover all 11 misleaders because each item in the bank is associated to only one misleader. The original \textsc{calvi} has 15 trick items, so we selected 15 as an upper bound.}

\begin{figure}[t]
  \centering 
  \includegraphics[width=\columnwidth]{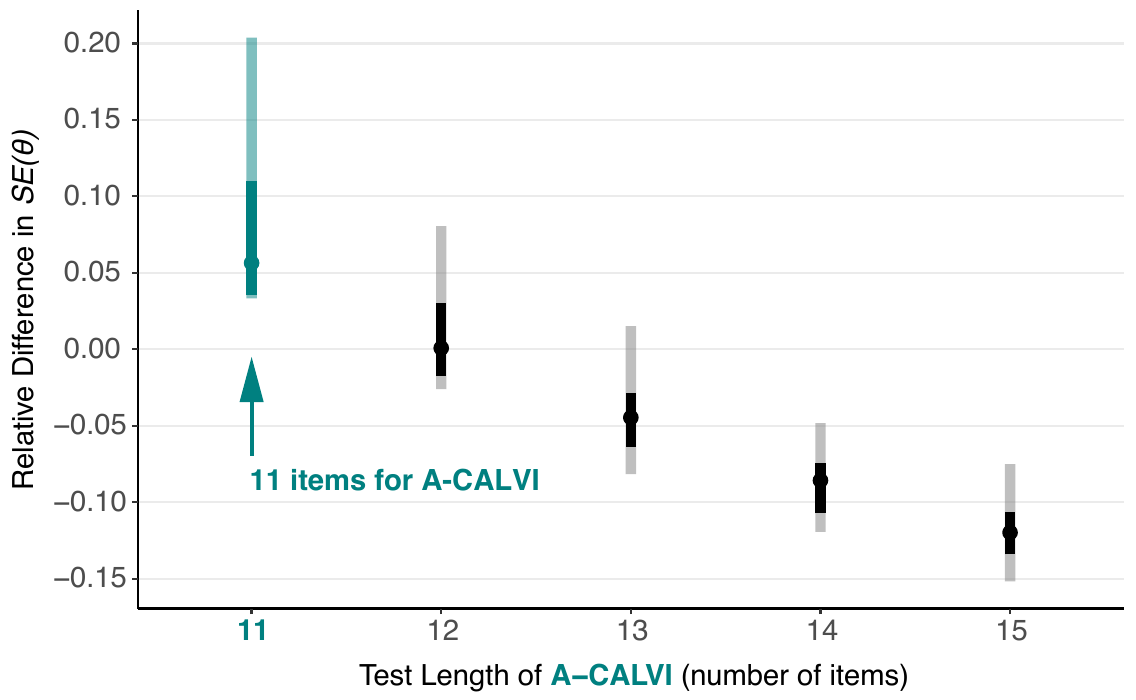}
    \vspace*{2pt}
  \caption{%
  	The distribution of $\textrm{relative difference in }SE(\theta)$ from length 11 to 15 for \ACALVI{a-calvi}, which shows a downward trend. \edit{We select 11 as the length of \ACALVI{a-calvi}. 
   }
  }
    \vspace*{-10pt}
  \label{fig:simresultscalvi}
\end{figure}

\paragraph{Results} \Cref{fig:simresultscalvi} shows a downward trend. From length of 13, the median and most observations of $\textrm{relative difference in }SE(\theta)$ fall below 0, indicating that \ACALVI{a-calvi} outperforms the original \textsc{calvi} in terms of measurement precision; this is because \ACALVI{a-calvi} selects better items from the bank of 45 than the fixed 15 items in \textsc{calvi}.
\edit{
 We select 11 as the length of \ACALVI{a-calvi} because it is shortest length that can cover all non-psychomtric features and the median and most of the observations of $\textrm{relative difference in }SE(\theta)$ at length 11 are below 0.10.
}

\subsection{Composition of \ACALVI{A-CALVI} Items via Qualitative Analysis}
\label{subsec:normal_items_study}

The developers of \textsc{calvi} included 15 items with well-constructed non-misleading visualizations (i.e. normal items) to accompany 15 items with misleading visualizations (i.e. trick items). This is because (1) if a test taker realizes that all items are about misleading visualizations, they could adopt the strategy of never selecting the obviously correct answers and (2) when people encounter visualizations in the real world,
they see both normal, well-constructed and misleading visualizations, which makes distinguishing misleading ones difficult \cite{CALVI}. However, a test taker's performance on the 15 normal items does not affect the $\theta$ estimate in \textsc{calvi}, and the normal items constitute half of the test, making it long and time-consuming. Therefore, we conduct a qualitative study to investigate the effect of normal items and whether they would be necessary in \ACALVI{a-calvi}.




We recruited two groups of participants from \edit{\code{\textbf{Prolific}}} for two conditions: (1) in \textit{without-normal condition}, participants \edit{took} \ACALVI{a-calvi} without normal items (i.e., only 11 adaptively-selected trick items from the bank of 45 trick items); (2) in \textit{with-normal condition}, they \edit{took} \ACALVI{a-calvi} with normal items (i.e., 11 adaptively-selected trick items + 11 fixed normal items). \edit{The 11 normal items were randomly sampled from the set of 15 normal items in the item bank of \textsc{calvi}.
}

After participants \edit{in both conditions completed the items on the test}, we asked them to answer a post-test survey of Yes/No and open-ended questions\footnote{The open-ended questions can be found in supplemental materials.} to investigate the possible impact of the presence of normal items. 
\edit{These questions included whether they noticed items with misleading charts (i.e., trick items) and how they approached the rest of the test after noticing trick items. We also provided the participants with the items they saw and their responses (presented in the order in their test) to help them recall the test content.} In the \textit{with-normal} condition, we also explore\edit{d} how participants approached the rest of the test after noticing the presence of normal items. 


\paragraph{Participants} 
A total of 60 participants were recruited
from \edit{\code{\textbf{Prolific}}} for this qualitative study, with 30 for the \textit{without-normal condition} and 30 for the \textit{with-normal condition}. The data of one participant in the \textit{with-normal condition} was discarded due to a data collection error, leaving a total of 59 participants. 
All participants fall under the following criteria: speak fluent English, have normal or corrected-to-normal vision, are between 18 and 65 years old, and are U.S. residents. Participants who failed more than half of the attention check questions were excluded\edit{, and all} participants passed the attention check in this study. \edit{In addition, \code{\textbf{Prolific}} participants from similar studies that the authors conducted in the past were excluded.}

\paragraph{Procedure} We started by presenting the consent form and information page describing the structure of the study to the participants. We instructed participants that they need to select an answer for the current item/question before proceeding to the next one, and once they moved on, they could not return to previous ones. \edit{After the participants completed the test items, they were directed to the post-test survey.} 

\paragraph{Method} We reviewed and coded participants' responses on the post-test survey to understand their experience with the test and how trick and normal items affected their approach to the test. 

\paragraph{Results} 
In the \textit{without-normal condition}, 24 out of 30 participants reported they noticed the presence of trick items, after which 16 paid more attention or became more skeptical of the items, 6 did not change how they approached the rest of the items, and 2 gave unclear responses.

In the \textit{with-normal condition}, 19 out of 29 participants reported that they noticed the presence of trick items, after which 14 of them paid more attention or double checked their answers, 4 did not change how they approached the rest of the items, and 1 participant gave an unclear response. When asked whether the presence of normal items changed their approach to answer trick items, 16 out of the 19 participants said no, and the remaining 3 participants mentioned that it only made them more suspicious of the normal items. 

There seems to be little evidence that the presence of normal items affected how participants approached the trick items, but we acknowledge that this sample size is small and the possibility of normal items having an effect still exists. Thus, we refrain from completely removing normal items, but instead look to reduce the amount. 

Our underlying goal is the same as \textsc{calvi}---the test taker sees a mix of trick and normal items, and we reason about this from the the test takers' perspective. Based on the analysis of trick items in the original \textsc{calvi}, there are certain misleaders that are substantially harder to identify,
such as \textit{Overplotting} or \textit{Missing Data} \cite{CALVI}, meaning these items are essentially perceived as ``normal'' in the eye of the test taker. Therefore, the items from such misleader categories can be effectively considered as ``normal'' items by the test takers (i.e., perceived ``normal''). We also notice that it is highly likely that the option of ``Cannot be inferred/inadequate information'' is the correct answer when it appears in a trick item in \ACALVI{a-calvi}. To prevent the strategy of always selecting this option when it appears, we decide to include in \ACALVI{a-calvi} 
all of the 
normal items
from \textsc{calvi} that contain this option but it is not the correct answer: there are 4 such items. 
With the 4 \textit{actual}
normal items and the \textit{perceived} ``normal'' items, we expect that, from the test taker's perspective, there would be a balanced mix of trick and ``normal'' items, following a similar format as the original \textsc{calvi}. Thus, \ACALVI{a-calvi} consists of 11 trick items selected adaptively by the \textsc{cat} algorithm and 4 \textit{actual} normal items. The total length of \ACALVI{a-calvi} is 15, half of the length of the original \textsc{calvi}.


\subsection{Final Configurations of \AVLAT{A-VLAT} and \ACALVI{A-CALVI}}\label{subsec:catfinalconfig}
\paragraph{\AVLAT{a-vlat}} 
The final configuration initializes the first item by using the mean of $\theta$'s from the \textsc{vlat} test tryout data to select the item with highest information in the bank. After the test taker answers each item, that item is removed from the bank, and $\theta$ is updated accordingly, which is then used to select the next item. Each time, \AVLAT{a-vlat} selects the next item with the highest information under the constraint that the 12 chart types and 8 tasks must be covered before the test terminates. \AVLAT{a-vlat} terminates after a test taker has answered 27 items. 

\paragraph{\ACALVI{a-calvi}}
The final configuration initializes the first item using the mean of $\theta$'s from the \textsc{calvi} test tryout data to select the item with highest information in the bank. After the test taker answers each item, that item is removed from the bank, and $\theta$ is updated accordingly, which is then used to select the next item. Each time, \ACALVI{a-calvi} selects the next item with the highest information under the constraint that the 11 misleaders must be covered before the test terminates. The 4 normal items are positioned at 4 randomly pre-determined locations among the trick items, but which normal item appears at which of the 4 positions is randomly determined at runtime for each test taker. \ACALVI{a-calvi} terminates after a test taker has answered 11 trick items and 4 normal items.


With the configurations of \AVLAT{a-vlat} and \ACALVI{a-calvi} finalized, it is necessary to know whether the two tests can provide stable and consistent measurements and whether they measure the abilities they are designed to measure. Therefore, we conduct four online studies to evaluate the reliability (\cref{sec:reliability}) and validity (\cref{sec:validity}) of the adaptive tests. 





\section{Reliability Evaluation}\label{sec:reliability}
Reliability of a test is whether
it can provide consistent and stable measurement. 
We provide evidence of \textbf{test-retest reliability}, a common way of assessing reliability and used in various domains to evaluate the reliability of \textsc{cat}s (e.g., \cite{Guinart2020CAT, Beiser2016testretest}).


Test-retest reliability evidence is gathered by administering a test to the same group of people twice with a time interval between the first and the second administrations and comparing participant's performances on each attempt. An interval of 1 or 2 weeks is typical \cite{polit2014getting}. For both \AVLAT{a-vlat} and \ACALVI{a-calvi}, we have the same study design: 
we recruit one group of participants from \code{Prolific}. Everyone takes the adaptive test and is invited back to take it again after a week.

We use the \textbf{intraclass correlation coefficient (\textsc{icc})} as the reliability metric \edit{recommended by} literature \cite{polit2014getting}. Next, we describe the models to compute \textsc{icc} for \AVLAT{a-vlat} and \ACALVI{a-calvi} and present the results.

\subsection{Model for Reliability Metric}\label{subsec:reliabilitymodel}
\textsc{icc} can be computed via a random effects model, and we used a Bayesian measurement error version of the model because the measurement for each person is a distribution rather than a point estimate. It is important to take into account the measurement error when computing \textsc{icc}, as otherwise it will be underestimated~\cite{wilms2020we}. Our model is preregistered on OSF (see link in abstract) and can be found in supplemental materials.\footnote{We initially pre-registered an implementation of this model that is difficult to fit and computes extra parameters (the latent $\theta^*$s) that we do not need. We later found an alternative parameterization better suited for our purpose, so the results in this section are from the alternative parameterization.} The model is:
\begin{align*}
    \theta^* &\sim \textrm{Normal}(\mu + \alpha_j, \sigma_{\epsilon}) \\
    \theta &\sim \textrm{Normal}(\theta^*, \textrm{se}(\theta)) \\
    \alpha_j &\sim \textrm{Normal}(0, \sigma_{\alpha}),
\end{align*}



\noindent where $\theta^*$ is the latent ability, $\mu$ is the unobserved mean, $\alpha_j$ is the random effect for person $j$, $\sigma_\epsilon$ is the standard deviation within-person, $\sigma_\alpha$ is the standard deviation between people, $\theta$ is the observed ability, and \textrm{se}($\theta$) is the measurement error of $\theta$. 
For the priors of this model, we selected $\textrm{Normal}(0,1)$ and $\textrm{Normal}(-1,1)$ for $\mu$ for \AVLAT{a-vlat} and \ACALVI{a-calvi}, respectively, because they are approximately the distributions of the estimated abilities in the \textsc{vlat} and \textsc{calvi} test tryout data. We also selected $\textrm{Normal}^+(0,1)$ for $\sigma_\alpha$ and $\sigma_\epsilon$.

For both the reliability of \AVLAT{a-vlat} and \ACALVI{a-calvi}, we ran our model with 4 chains, each with 20,000 iterations. We discarded 10,000 warmup iterations per chain and thinned the final sample by 5, yielding 8,000 total post-warmup draws.

This model outputs distributions of $\sigma_\alpha$ and $\sigma_{\epsilon}$, which we then use to compute \textsc{icc} with the following formula \cite{liljequist2019intraclass}: $$\text{\textsc{icc}} = \frac{\sigma_{\alpha}^2}{\sigma_{\alpha}^2+\sigma_{\epsilon}^2}.$$ \textsc{icc} can be understood as the ratio between the variance between people ($\sigma_{\alpha}^2$) and the sum of variance between people and the (undesirable) within-person variance ($\sigma_{\epsilon}^2$) \cite{liljequist2019intraclass}. 





\subsection{Reliability of \AVLAT{A-VLAT}}\label{subsec:study3}
\paragraph{Participants} 90 participants were recruited in anticipation that around 60 would return to the second study so that we have a sample size close to 60. This sample size was determined by collecting pilot data with 20 participants, fitting the model with pilot data, generating simulated participants with the fitted model, and choosing the sample size that would yield a \textpm0.1 95\% credible interval around the \textsc{icc}.\footnote{\edit{All pilot analyses} for sample size determination are in supplemental materials\looseness=-1.} 



In the end, 47 participants
completed both parts of the study. All participants fall under the same criteria listed in \cref{subsec:normal_items_study}.

    \paragraph{Model Diagnostics}
The minimal bulk effective and the minimum tail effective sample sizes are 6,757 and 7,067, and the $\Hat{R}$ values are approximately 1.00 for all parameters. 



\begin{figure}[t]
  \centering  \includegraphics[width=\columnwidth
  ]{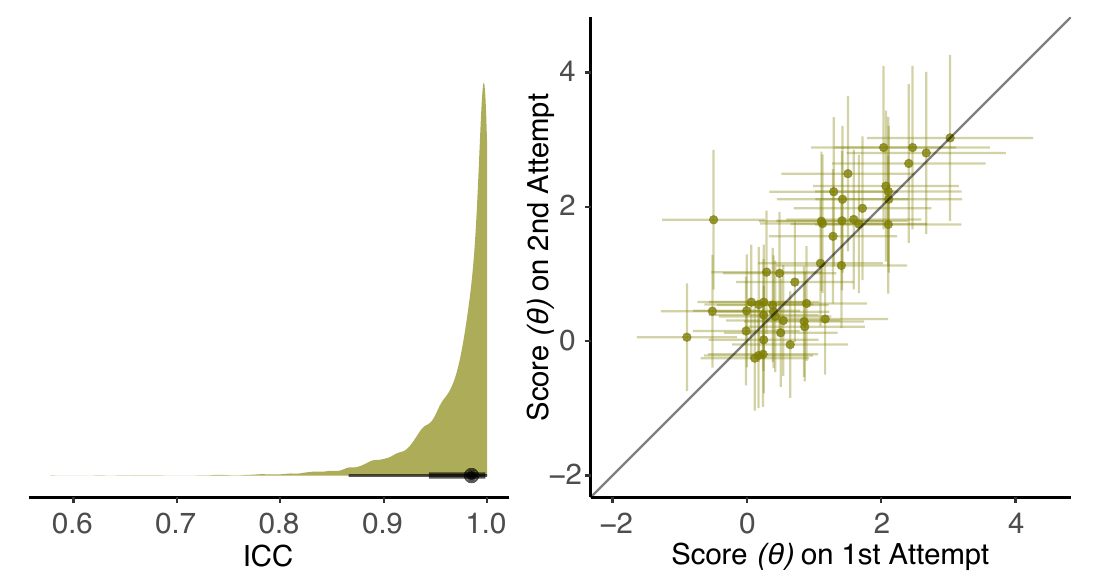}
      \vspace*{3pt}
  \caption{Reliability of \AVLAT{a-vlat}: the left plot is a posterior density plot of \textsc{icc} with a point interval plot at the bottom; the point is the median, and the lighter and darker bars are the 95\% and 66\% intervals.
  The right plot is a scatterplot of the 47 participants' scores on the first and second attempts, where each point represents a participant, and the cross lines at each point represents the measurement error (95\% intervals).
  }
    \vspace*{-12pt}
\label{fig:reliability_vlat}
\end{figure}

\paragraph{Results}
\Cref{fig:reliability_vlat} shows the posterior density of \textsc{icc} and a scatterplot of participants' scores on the first attempt versus those on the second attempt. \textbf{The median \textsc{icc} is 0.98 with 95\% CI of [0.87, 1.00]}, so \AVLAT{a-vlat} has \textbf{excellent test-retest reliability} \cite{cicchetti1994guidelines, matheson2019we}. 

\subsection{Reliability of \ACALVI{A-CALVI}}\label{subsec:study4}
\paragraph{Participants} We determined the sample size by first following the same method in \cref{subsec:study3}. However, in the model with measurement error on pilot data, the intervals of \textsc{icc} did not shrink as the sample size of the simulated participants increased. Therefore, we use a sample size of 60 because it is consistent with the sample size of the reliability study for \AVLAT{a-vlat} and the 95\% intervals on $\sigma_\alpha$ and $\sigma_\epsilon$ have similar width as those in \cref{subsec:study3}. In addition, we report the \textsc{icc} results from models with and without measurement error because the interval of \textsc{icc} may be large for the model with measurement error. 
Based on the attrition rate of the previous study, we recruited 115 participants, in anticipation that around 60 would return to the second study. In the end, 73 participants
\edit{from} the first part of the study returned to the second part. All participants fall under the same criteria listed in \cref{subsec:normal_items_study}.

\paragraph{Model Diagnostics}
The minimal bulk effective and the minimum tail effective sample sizes are 7,272 and 7,560, and the $\Hat{R}$ values are approximately 1.00 for all parameters.

\begin{figure}[t]
  \centering  \includegraphics[width=\columnwidth
  ]{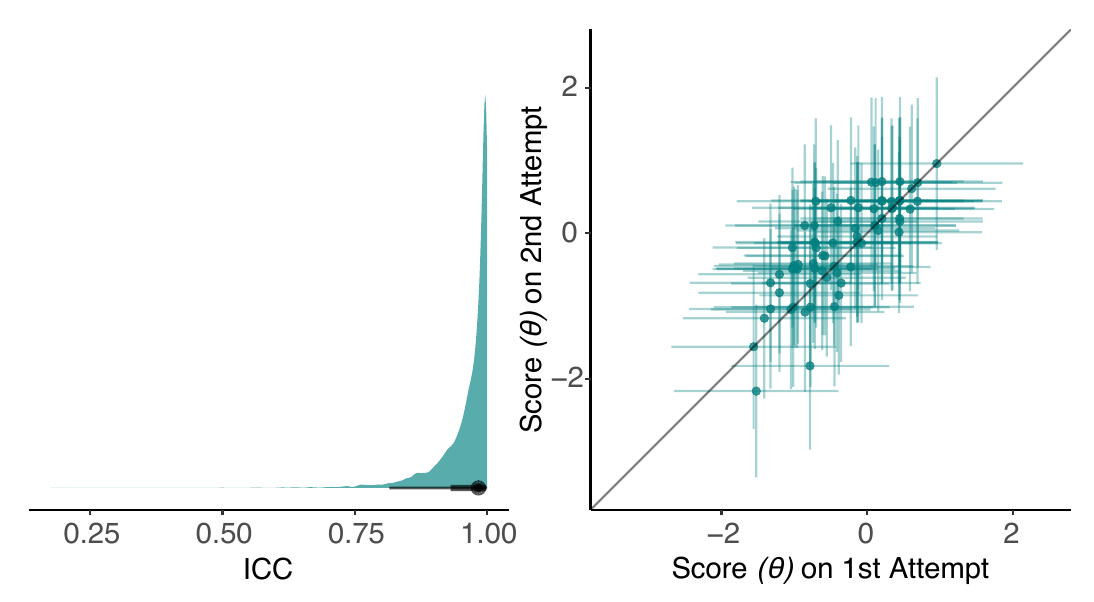}
    \vspace*{-10pt}
  \caption{Reliability of \ACALVI{a-calvi}: the left plot is a posterior density plot of \textsc{icc}; the right plot is a scatterplot of the 73 participants' scores on the first and second attempts with measurement error.
  }
    \vspace*{-10pt}
\label{fig:reliability_calvi}
\end{figure}

\paragraph{Results}
\Cref{fig:reliability_calvi} shows the posterior density of \textsc{icc} and a scatterplot of participants' scores on the first attempt versus those on the second attempt. \textbf{The median \textsc{icc} is 0.98 with 95\% CI of [0.82, 1.00]}, so \ACALVI{a-calvi} has \textbf{excellent test-retest reliability} \cite{cicchetti1994guidelines, matheson2019we}. 
Even though the model with measurement error is able to produce a small interval for \textsc{icc} with data from this study, we report the result from the other model as stated in pre-registration: the median \textsc{icc} is 0.77 with 95\% CI of [0.66, 0.85]. We select the \textsc{icc} from the model with measurement error as the final report for reliability because a model that does not account for measurement error consistently underestimates \textsc{icc}. \cite{wilms2020we}. 





\section{Validity Evaluation}\label{sec:validity}
We evaluate whether our adaptive tests are valid---that they measure what they are designed to measure---by computing the \textbf{convergent validity} \cite{Chin2014ConvergentValidity}: the correlation between test takers' performance on the adaptive tests and their performance on the original static tests. This is an appropriate measure of the validity of the adaptive tests \cite{green1984technical} because \textsc{vlat} and \textsc{calvi} are both valid measurements \cite{CALVI, Lee2017VLAT} and the equivalence of the original and adaptive tests through strong correlation would demonstrate that the adaptive tests indeed measure the same abilities as the originals, hence are valid.





To compute the correlation coefficient for convergent validity, the same group of participants need to take both versions of the tests. Because retaking the tests may create learning and ordering effects, we use two groups of participants to counterbalance the effects.
For both \AVLAT{a-vlat} and \ACALVI{a-calvi}, we have the same study design with two groups, both recruited from \code{Prolific}: each person takes both tests, with a week\footnote{Because of the similarity of content between the adaptive and original tests, we choose to have a one-week interval to mitigate learning effects.} in between sessions. Order is counterbalanced: half the participants (Group 1) takes the original test first, and the other half (Group 2) takes the adaptive test first.
The procedure of our validity studies is the same as that in \cref{subsec:normal_items_study}.


\subsection{Model for Validity Metric} 
Correlation coefficients can be computed via a Bayesian bivariate Normal model.
We again use a measurement error model. The model is preregistered on OSF (see link in abstract) and can be found in supplemental materials.\footnote{The model results in this section are based on an alternative implementation different than \edit{the} initially pre-registered for the same \edit{reason in \cref{subsec:reliabilitymodel}.}} 
The bivariate Normal model is:
\begin{align*}
    \begin{bmatrix} \theta_{\texttt{O}}^* \\ \theta_{\texttt{A}}^* \end{bmatrix} &\sim \textrm{Normal}_{2} \left(\begin{bmatrix} \mu_{\texttt{O}} \\ \mu_{\texttt{A}} \end{bmatrix}, \sigma_{\texttt{O}},
    \sigma_{\texttt{A}},
    \rho \right) \\
    \theta_{\texttt{O}} &\sim \textrm{Normal}(\theta_{\texttt{O}}^*, \textrm{se}(\theta_{\texttt{O}})) \\
    \theta_{\texttt{A}} &\sim \textrm{Normal}(\theta_{\texttt{A}}^*, \textrm{se}(\theta_{\texttt{A}})).
\end{align*}
\noindent where $\theta_{\texttt{O}}^*$ and $\theta_{\texttt{A}}^*$ are the latent abilities measured from the original test and the adaptive test, with $\mu_{\texttt{O}}$ and $\mu_{\texttt{A}}$ as their means, $\sigma_{\texttt{O}}$ and $\sigma_{\texttt{A}}$ as their standard deviations, and $\rho$ as the correlation between them. $\theta_{\texttt{O}}$ and $\theta_{\texttt{A}}$ are the observed abilities, with measurement errors $\textrm{se}(\theta_{\texttt{O}})$ and $\textrm{se}(\theta_{\texttt{A}})$. In the implementation of this model, we re-parameterized $\mu_{\texttt{O}}$ and $\mu_{\texttt{A}}$ in terms of the difference between them by subtracting the mean of $\theta_{\texttt{O}}$ from all the observed $\theta$'s.
For the prior of the model, we selected $\textrm{Normal}(0,1)$ for the difference between $\theta_{\texttt{O}}$ and $\theta_{\texttt{A}}$, as 1 is a large difference in latent \textsc{irt} ability. 
We picked $\textrm{Normal}(1,0.5)$ for $\sigma_{\texttt{O}}$ and $\sigma_{\texttt{A}}$, because standard deviations should center around 1 as the \textsc{irt} model is parametrized so that latent abilities have a standard deviation around 1. For $\rho$, we selected a uniform prior (the LKJ distribution with parameter $\eta = 1$).\footnote{We pre-registered LKJ distribution with parameter $\eta = 3$ as the prior of $\rho$ in the initial implementation of the model for \AVLAT{a-vlat}, because it needed a tight prior to converge. However, the alternative implementation---which does not estimate all $\theta^*$s---converges much more easily, so we relaxed the prior on $\rho$.} 

We ran our final model with 4 chains, each with 20,000 iterations. We discarded 10,000 warmup iterations per chain and thinned the final sample by 5, yielding 8,000 total post-warmup draws.

This model outputs the distribution of the correlation coefficient $\rho$ for convergent validity. 





\subsection{Validity of \AVLAT{A-VLAT}}\label{subsec:study1}
\label{subsec:vlat_cat_validity}
\paragraph{Participants}
120 participants were recruited (60 for each group) in anticipation around 100 would return to the second part of the study so that we have a sample size close to 100. This sample size was determined by collecting pilot data with 10 participants in each group, fitting the model with pilot data, generating simulated participants with the fitted model, and choosing the sample size that would yield a \textpm0.1 95\% credible interval around $\rho$.

In the second part of the study, 49 returned in Group 1 and 37 returned in Group 2, for a total of 86 participants.
All participants fall under the same criteria listed in \cref{subsec:normal_items_study}. 

\paragraph{Model Diagnostics}
The minimal bulk effective and the minimum tail effective sample sizes are 7,607 and 7,394, and the $\Hat{R}$ values are approximately 1.00 for all parameters.

\begin{figure}[t]
  \centering
  \includegraphics[width=\columnwidth
  ]{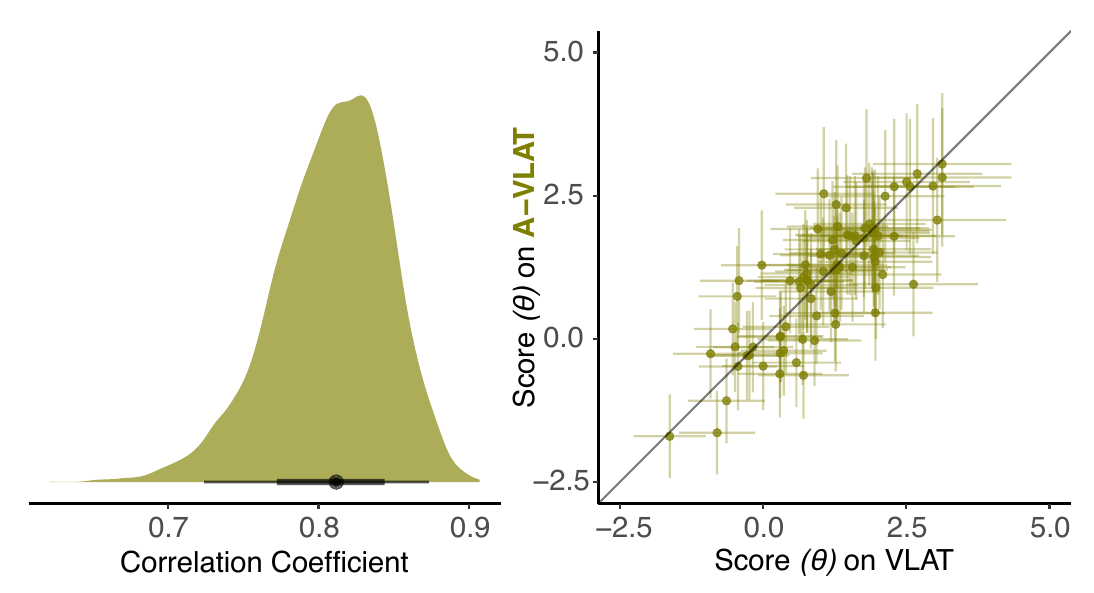}
    \vspace*{-5pt}
  \caption{Validity of \AVLAT{a-vlat}: the left plot is a posterior density plot of correlation coefficient $\rho$; the right plot is a scatterplot of the 86 participants' scores on \textsc{vlat} and \AVLAT{a-vlat} with measurement error.}
    \vspace*{-6pt}
\label{fig:validity_vlat}
\end{figure}

\paragraph{Results}
\Cref{fig:validity_vlat} shows the posterior density of $\rho$ and a scatterplot of participants' scores on the adaptive test versus those on the original test. \textbf{The median $\rho$ is 0.81 with 95\% CI of [0.72, 0.87]}. While there is no gold standard for convergent validity, a correlation coefficient below $0.50$ should be avoided and above $0.70$ is recommended \cite{carlson2012validitythreshold}, so \AVLAT{a-vlat} has \textbf{high convergent validity}. 

\subsection{Validity of \ACALVI{A-CALVI}}\label{subsec:study2}
\paragraph{Participants} 
A sample size of 100 was determined in the same way as in \cref{subsec:study1}. Based on the attrition rate of the previous study, 140 participants were recruited (70 in each group), in anticipation that around 100 participants would return to the second part of the study.



In the first part of the study, one participant's data was discarded because they failed the attention check questions. In the second part, 47 and 43 returned in Groups 1 and 2, for a total of 90 participants.
All participants fall under the same criteria listed in \cref{subsec:normal_items_study}.


\paragraph{Model Diagnostics}
The minimal bulk effective and the minimum tail effective sample sizes are 6,455 and 5,090, and the $\Hat{R}$ values are approximately 1.00 for all parameters.

\begin{figure}[t]
  \centering
  \includegraphics[width=\columnwidth
  ]{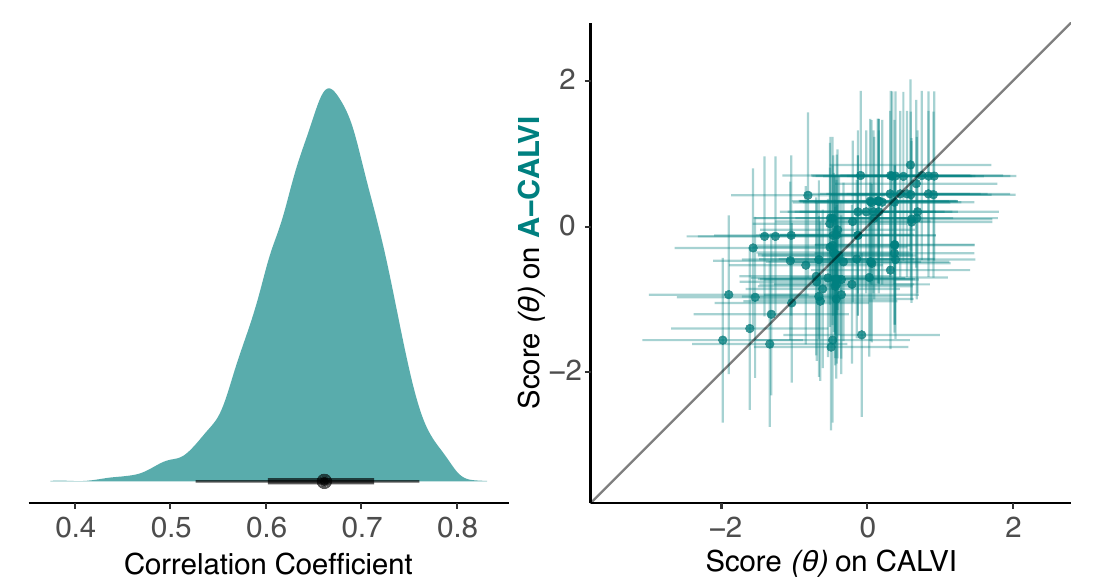}
   \vspace*{-5pt}
   \caption{Validity of \ACALVI{a-calvi}: the left plot is a posterior density plot of the correlation coefficient $\rho$; the right plot is a scatterplot of the 90 participants' scores on \textsc{calvi} and \ACALVI{a-calvi} with measurement error.}
    \vspace*{-7pt}
\label{fig:validity_calvi}
\end{figure}

\paragraph{Results}
\Cref{fig:validity_calvi} shows the posterior density of $\rho$ and a scatterplot of participants' scores on the adaptive test versus those on the original test. \textbf{The median $\rho$ is 0.66 with 95\% CI of [0.53, 0.76]}. Since a correlation coefficient below $0.50$ should be avoided and above $0.70$ is recommended \cite{carlson2012validitythreshold},  \ACALVI{a-calvi} has \textbf{acceptable convergent validity}.

\section{Discussion}

\subsection{How Reliable, Valid, and Precise  Are \textsc{cat}s Really?
}
The results from \cref{sec:reliability} and \cref{sec:validity} have shown that we have successfully created two reliable and valid adaptive tests, \AVLAT{a-vlat} and \ACALVI{a-calvi}, which measure the same abilities as the original static \textsc{vlat} and \textsc{calvi}. In addition, we collected data that can be used to demonstrate the measurement precision of both adaptive tests:
in validity evaluation, every participant was asked to take both the adaptive and original versions. We can use this data to analyze the $\textrm{relative difference in }SE(\theta)$ of both adaptive tests as we did in \cref{subsec:termination} with simulation data.\footnote{The code for this analysis is provided in supplemental materials.}

For \AVLAT{a-vlat}, the distribution of $\textrm{relative difference in }SE(\theta)$ for 86 participants has a mean of 0.089 (i.e., for a person with average ability in this group, the standard error for them is within 8.9\% of the standard error of the baseline.) with \edit{a} standard deviation \edit{of} 0.11. 

For \ACALVI{a-calvi}, the distribution of $\textrm{relative difference in }SE(\theta)$ for 90 participants has a mean of 0.023 with \edit{a} standard deviation \edit{of} 0.039.

We observe that results of $\textrm{relative difference in }SE(\theta)$ from actual participants reveal that both adaptive tests are similarly precise compared to their original static counterparts, in accordance with the simulation results 
in \cref{subsubsec: vlatcatsim} and \cref{subsubsec:simcalvi}. This demonstrates the measurement precision of \AVLAT{a-vlat} and \ACALVI{a-calvi} and further supports the justification for the shortened lengths of the adaptive tests. 

\edit{Another consideration of reliability is whether a test taker can recover quickly if they mistakenly select a wrong answer. First, we define $d_i$ as the absolute value of the difference between the mean of the $\theta$ estimate distribution and the true $\theta$ of a test taker after answering the $i$-th item in the test. Then, ``mistakenly selecting a wrong answer for item $i$'' is defined as when $d_i > d_{i-1}$; i.e., when their $\theta$ estimate moves away from the true $\theta$ after answering an item. Naturally, ``recovery'' from choosing a wrong answer for item $i$ happens at the first step $i^\prime$ where $d_{i^\prime} \le d_i$; i.e., their $\theta$ estimate moves back to being within at least $d_i$ of their true $\theta$, and recovery length is $i^\prime - i$. We conducted simulations\footnote{\edit{The code for the simulations is provided in supplemental materials.}} to study the recovery lengths in our adaptive tests. We generated 500 simulated participants for each test as before, and simulated their response correctness to \AVLAT{a-vlat} and \ACALVI{a-calvi}. The median recovery lengths for \AVLAT{a-vlat} and \ACALVI{a-calvi} are 3 and 2 with standard deviations of 4.8 and 1.1, respectively. Therefore, test takers can recover quickly from a mistake in our tests, demonstrating good reliability. 
}

\subsection{Correlation within Chart Types, Tasks, and Misleaders} \label{subsec:correlation_discussion}
Items in the banks of \textsc{vlat} and \textsc{calvi} span a range of chart types, tasks, and misleaders. Studying the relationship between people's performance within these different categories of visualization features can shed light on how different aspects and skills of visualization literacy connect and transfer, which can then help us better understand this ability and potentially further optimize adaptive testing. 

Actual participants have been asked to answer these visualization literacy items from prior work \cite{CALVI, Lee2017VLAT}, 
as well as the qualitative study, reliability evaluation, and validity evaluation in this paper. This gives us the opportunity to use this data to examine the relationship between participants' performance on different chart types, tasks, and misleaders. Therefore, we conducted an exploratory study to compute the correlations within these visualization features by using both test tryout data (\textsc{vlat} and \textsc{calvi}) and data collected in this paper and employing the same \textsc{irt} models described in \cref{sec:itemanalysis}.\footnote{The entire results can be found in supplemental materials.} Next, we present some highlights from this exploration and discuss their implications. 

\paragraph{\textsc{VLAT} Chart Types}
We found high correlations between performance on certain chart types: 0.83 with 95\% CI of [0.72, 0.92] for scatterplot and bubble chart, 0.82 with 95\% CI of [0.72, 0.90] for 100\% stacked bar and stacked bar charts, and 0.81 with 95\% CI of [0.69, 0.90] for scatterplot and line chart. There are also chart types without high correlations with others: choropleth map's and stacked area chart's correlation with other chart types do not exceed 0.38 and 0.52, respectively. 

\paragraph{\textsc{VLAT} Tasks}
Many pairs of tasks have high correlations: 0.93 with 95\% CI of [0.88, 0.97] for \textit{retrieve value} and \textit{make comparisons}, 0.89 with 95\% CI of [0.81, 0.96] for \textit{find extremum} and \textit{make comparisons},  0.79 with 95\% CI of [0.70, 0.86] for \textit{determine range} and \textit{retrieve value}. Most tasks highly correlate with at least another task.

\paragraph{\textsc{CALVI} Misleaders}
Most pairs of misleaders have weak or moderate positive correlations: \textit{Manipulation of Scales -
Inappropriate Scale Range} and \textit{Manipulation of Scales -
Inappropriate Use of Scale Functions} have a correlation of 0.75 with 95\% CI of [0.58, 0.89], highest among all pairs of misleaders. Some misleaders have very weak correlations with others; \textit{Overplotting} especially stands out as its correlations with the other misleaders range from -0.21 to 0.34. This is reasonable as \textit{Overplotting} is one of the most difficult misleaders to judge correctly. 

Although \AVLAT{a-vlat} and \ACALVI{a-calvi} are significantly shorter than their original versions, there are opportunities to further shorten them by leveraging these correlations. For instance, the \textsc{vlat} tasks \textit{retrieve value} and \textit{make comparisons} are highly correlated, so test developers could make the item selection process less constrained by only covering one of the two instead of both. Further research on incorporating these correlations into \textsc{cat} algorithms for visualization literacy is needed. 

Moreover, these correlations are valuable quantitative data that informs taxonomy in visualization literacy: people's performance on tasks \textit{retrieve value}, \textit{find extremum}, and \textit{make comparisons} are highly correlated, which means it may be reasonable to categorize them into one task. Additionally, when researchers and educators design curricula or tools to improve people's visualization literacy, they can draw insights from these correlations to create more efficient and effective interventions: since people's performances on answering items with scatterplots and bubble charts are strongly correlated, teachers may choose to focus on one, or group lessons together for increased efficiency\looseness=-1.  

\subsection{\edit{Customization of} Adaptive Tests}
\label{sec:cats:customization}
When using \AVLAT{a-vlat} and \ACALVI{a-calvi}, test administrators can customize many components of the tests, including initialization, item selection, and termination criterion. For initialization, we chose the means of the $\theta$ distributions from test tryouts. However, test administrators can explore other alternatives, such as randomly choosing the first item from the bank to decrease the item exposure (i.e., the frequency at which an item is administered across all administrations) of certain items. Test administrators can also customize the non-psychometric constraints in item selection. For example, if one wishes to focus on only covering all \textsc{vlat} chart types or only covering all tasks, they could adjust the non-psychometric constraints to achieve that. Moreover, the results from the simulation studies in \cref{subsec:termination} can guide administrators to adjust the length of the adaptive test based on their time-limit constraints and requirement for measurement precision. For scoring, we recommend reporting the mean of the posterior distribution of $\theta$ as a point estimate (described in \cref{subsec:scoring}), as well as the raw correctness (i.e., number of items answered correctly over the total number of items answered) to help test takers intuitively interpret their final scores. 

\edit{Test administrators may wonder about the reliability and validity of their customized tests.}
\edit{We think that validity would remain relatively stable as the lengths of adaptive tests change so long as the non-psychometric features are covered, because these features ensure that the adaptive and original versions are qualitatively similar. Also, we think that the more non-psychometric features not covered, the less valid the adaptive tests would likely be. However, if two features are very similar, as discussed in \cref{subsec:correlation_discussion}, not covering one may not have a noticeable effect on validity. In addition, test-retest reliability could vary if the length is too short or long. On the one hand, if too short, the $\theta$ estimates would be more susceptible to measurement error and have limited variability, which can affect reliability. On the other hand, if a test is too long, test takers' motivation, attention, and effort might change on the second attempt, causing the test-retest reliability to decrease. Although test length can affect reliability and validity in extreme cases, we believe that moderately changing the length of \AVLAT{a-vlat} or \ACALVI{a-calvi} would have a marginal effect on reliability and validity so long as the non-psychometric constraints are not violated. 
}

For test administrators who only need to measure the visualization literacy of a group of people once, they can use \AVLAT{a-vlat} and \ACALVI{a-calvi} as alternatives to their original versions because they are much shorter, contain balanced content, and have similar measurement precision. For researchers and educators who want to conduct pre-post testing to evaluate the effectiveness of their interventions by administering a visualization literacy test once before and once after the intervention, we recommend using our adaptive tests for both administrations. For the second administration, they can choose to exclude certain overused items in the first attempt if they are concerned about item exposure. They could also use the estimated $\theta$ of a particular person in the first attempt to initialize this person's first item in the second attempt. 



\subsection{Limitation and Applicability in Practice} 

\edit{To create assessments under the \textsc{irt} framework, it is necessary to construct items, run a test tryout study, and compute item parameters using \textsc{irt} models. This process requires substantial resources as well as technical expertise on \textsc{irt} modeling, which can be a barrier to implementing adaptive tests in practice. Moreover, adaptive assessments} require more complex infrastructure to deliver than a standard series of questions and it is not trivial to add newly designed test items, due to the adaptive nature requiring computation at each step.
 However, given the benefits of \textsc{cat}s, there is a need to democratize community access to capabilities for test development and administration.
Both \edit{\AVLAT{a-vlat} and \ACALVI{a-calvi}} are therefore implemented using an experimental framework, designed to create flexible and reusable visualization assessments. 
One key feature is the decoupling of static and dynamic experiment/assessment content, which supports the implementation of \textsc{cat}s in a more generalized fashion.
Consequently, \AVLAT{a-vlat} and \ACALVI{a-calvi} assessments are designed such that they can be reshaped by intermediate programmers, making it easier for researchers and practitioners to adapt these assessments for specific needs. 
As more assessments are developed in the visualization community, such infrastructure could play a role in developing adaptive versions of future tests, while also providing helpful validity, reliability, and general test analysis capabilities. Code is available in the paper's supplemental \edit{materials}.

\subsection{Automated Item Generation}
Although \AVLAT{a-vlat} and \ACALVI{a-calvi} are half the lengths of their original versions, the size of the item banks are still limited, meaning that repeated administrations of the tests would increase the item exposure quickly. Larger banks can reduce item exposure and allow for more attempts of re-assessments without giving out the same items. However, generating large banks is costly, both in terms of time and resources. The current process is not scalable because it requires manually creating the visualizations and writing the question text for each item, as well as recruiting a large number of participants for test tryout in order to estimate the item parameters needed for adaptive testing.

There is an opportunity for automating the item generation process and inferring item parameters without recruiting a large number of actual participants: the items in the banks are made up of specific components that are common in all items, such as the visualizations, question text structure, and the answer options. One can potentially leverage visualization grammars and machine learning techniques to automatically generate items and infer their item parameters using substantial existing data. Future studies can investigate ways to scale and improve the item generation process. 

\section{Conclusion}
To more efficiently measure visualization literacy, we develop \AVLAT{a-vlat} and \ACALVI{a-calvi}, which are adaptive versions of \textsc{vlat} and \textsc{calvi} that are half the length of the original tests. We conduct a qualitative study to improve the question composition of \textsc{calvi}, and incorporate non-psychometric features of both assessments as constraints, ensuring that the adaptive tests cover all tasks and chart types (for \textsc{vlat}) or misleaders (for \textsc{calvi}) from the original assessments. 
We establish the reliability and validity of these adaptive tests via four online studies and associated analyses. 
We demonstrate how to apply \textsc{cat} to create more efficient visualization literacy assessments that are reliable, valid, and precise. 
As visualization literacy can influence data-driven decisions, both \AVLAT{a-vlat} and \ACALVI{a-calvi} can be used to support the study of visualization literacy development and intervention evaluations through pre- and post-testing. 
Adaptive assessments can transform measurement practice in visualization literacy, and our work establishes a channel to support the adoption of more efficient assessments in practice.

\section*{Acknowledgments}
We extend our gratitude to William Revelle and Elizabeth Tipton for their early-stage feedback. We thank Mandi Cai, Maryam Hedayati, and members of the MU Collective at Northwestern for their support and feedback. We are grateful for the Design Cluster program at the Center for HCI and Design at Northwestern, especially Haoqi Zhang, Kapil Garg, and Eleanor O'Rourke for their mentorship. We thank Bum Chul Kwon and other developers of \textsc{vlat} for sharing their data. We are grateful for Matthew vonAllmen and others who volunteered their time in testing our system before the launch of our experiments. We also thank the online participants for their time and the reviewers for their insightful comments. This work was supported in part by grants from the National Science Foundation (\#1815587, \#2120750, \#2127309 to the Computing Research Association for the CIFellows Project).


\section*{Supplemental Materials}
\label{sec:supplemental_materials}


All supplemental materials are available on OSF at \url{https://osf.io/a6258/}.






	

\bibliographystyle{abbrv-doi-hyperref}

\bibliography{template}










\end{document}